\documentclass[aps,preprint]{revtex4}
\usepackage{amsfonts}
\usepackage{amsmath}
\usepackage{amssymb}
\usepackage{graphicx}

\input{tcilatex}

\begin{document}

\preprint{}
\title[Short title for running header]{Vortex driven phase transition in
Topologically Massive QED }
\author{Yuichi Hoshino}
\affiliation{Kushiro National College of Technology,Otanoshike Nishi-2-32-1,Kushiro
City,Hokkaido 084-0916,Japan}
\pacs{PACS number}

\begin{abstract}
There is chiral like symmetry for 4-component massless fermion in
(2+1)-dimensional gauge theory.Since QED$_{3}$ with Chern-Simons term
contains vortex solution for vector potential,one may expect vortex driven
phase transition as Kosterlitz-Thouless type where chiral condensate is
washed away at zero temperature.To study this possibility,we evaluate the
fermion propagator by Dyson-Schwinger equation numerically and spectral
function analytically in the Landau gauge.For quenched case we adopt
Ball-Chiu vertex to keep gauge invariance of the results.The critical value
of topological mass,above which chiral condensate washed away, turned out to
be $O(10^{-2})e^{2}$ at least for weak coupling in both cases.
\end{abstract}

\volumeyear{year}
\volumenumber{number}
\issuenumber{number}
\eid{identifier}
\startpage{101}
\endpage{102}
\maketitle
\tableofcontents

\section{ \quad\ Introduction}

Dynamical effects caused by Chern-Simons terms have been known in various
places.For examples it is well-known to explain the quantum Hall effects,so
called statistics transmutation\cite{POLYAKOV}.On the other hand vortex has
an important role to wash away long-range order in XY model or condensate of
(2+1)-dimensional model for superfluid helium film \cite{KOSTERLITZ:1973},%
\cite{KOSTERLITZ:1974}.These are known as Kosterlitz-Thouless transition.
However its detailed dynamics have not been known.It is an interesting
problem to study the dynamics of the above phase transition.In this sense QED%
$_{3}$ with Chern-Simons term gives us an example to show an dynamical
effects of vortex on chiral condensate. QED$_{3}$ with Chern-Simons term has
a vortex solution for vector potential by solving Maxwell equation with
charged particle \cite{DJT}. Therefore the situation is similar to the
isolated vortex inside superfluid at high temperature.We can examine the
phase transition by solving Dyson-Schwinger equation for the fermion
self-energy for chiral condensate and its destruction by vortices.In this
model Chern-Simons term is absorbed to parity odd part of the gauge boson
propagator and gauge boson acquires a mass.Pure QED part of gauge boson
contributes to the condensation of $e\overline{e}$,while the latter may wash
away the condensate at zero temperature.These are the main goals of our
analysises.In 4-dimensional representation of spinor we have chiral symmetry
for massless fermion.If it breaks dynamically we have two kinds of mass as
chiral symmetry breaking and parity violation.For infinitesimal value of the
topological mass it has been pointed out by K.I.Kondo and P.Maris that
chiral symmetry restores and parity violating phase remains within $1/N$
expansion and nonlocal gauge of Dyson-Schwinger equation \cite{Kondo},\cite%
{Raya},\cite{Hiroshima}.Here we consider weak coupling and gauge covariant
approximation which satisfy Ward-Takahashi relation at first.After that we
examine the $1/N$ expansion in the Landau gauge.In the final section we
evaluate the spectral function of the fermion propagator \cite{OPS},\cite{HOSHINO} .This method
is helpful to determine the infrared behaviour or one particle singularity
of the propagator in the existence of massless boson as photon,which has
been known in QED$_{3+1}$.Since QED$_{2+1}$ is super renormalizable we get a
short distance behaviour of the fermion propagator too.As a result we show
the effect of vortex on the lowest order spectral function of parity even
scalar part of the propagator.So that we find the critical value of
topological mass above which chiral condensate is washed away. Fortunately
this value coincides with that obtained by numerical analysis of
Dyson-Schwinger equations.

\section{\protect\bigskip Topologically massive QED}

The Lagrangian density of Topologically Massive QED$_{3}$ is written\cite%
{DJT}. 
\begin{equation}
\tciLaplace=\overline{\psi}(i\gamma\cdot(\partial-ieA)-m)\psi-\frac{1}{4}%
F_{\mu\nu}F^{\mu\nu}-\frac{\mu}{4}\epsilon^{\mu\nu\rho}F_{\mu\nu}A_{\rho}-%
\frac{1}{2\eta}(\partial\cdot A)^{2}.
\end{equation}
This system is characterized by equations of motion 
\begin{align}
(i\gamma\cdot(\partial-ieA)-m)\psi(x) & =0, \\
\partial_{\mu}F^{\mu v}+\frac{\mu}{2}\epsilon^{\nu\alpha\beta}F_{\alpha%
\beta} & =J^{\nu}.
\end{align}
Under the gauge transformation 
\begin{align}
\psi & \rightarrow e^{i\Omega}\psi, \\
A_{\mu} & \rightarrow A_{\mu}+\frac{1}{e}\partial_{\mu}\Omega,
\end{align}
Lagrangian is invariant but the gauge fixing term changes by a total
derivative%
\begin{equation}
\tciLaplace_{g}\rightarrow\tciLaplace_{g}+\partial_{\alpha}(\frac{\mu}{4e}%
\epsilon^{\alpha\mu\nu}F_{\mu\nu}\Omega).
\end{equation}
However action $\int d^{3}x\tciLaplace(x)$ is gauge invariant.It is easily
seen by partial integration.

\subsection{vortex solution}

There exists a vortex solution of vector potential in Topologically Massive
QED$_{3}$ \cite{DJT}.It has been known by solving equation of motion at
large $|x|$ 
\begin{equation}
\partial_{\mu}F^{\mu\nu}+\mu\epsilon^{\nu\alpha\beta}F_{\alpha\beta}=J^{\nu}.
\end{equation}
We separate the above equation into time and space components%
\begin{align}
\triangledown\cdot\mathbf{E}-\mu B & =\rho, \\
\epsilon^{ij}\partial_{j}B\mathbf{-}\frac{\partial E^{i}}{\partial t}\mathbf{%
+}\mu\epsilon^{ij}E^{j} & =J^{i}\mathbf{,}
\end{align}
where $B\mathbf{=-}1/2\epsilon^{ij}F_{ij}=\epsilon^{ij}%
\partial_{i}A_{j},E^{i}=F^{i0}$. Using dual field strength 
\begin{align}
^{\ast}F^{\mu} & =\frac{1}{2}\epsilon^{\mu\alpha\beta}F_{\alpha\beta}, \\
F^{\mu\nu} & =\epsilon^{\mu\nu\alpha}\text{ }^{\ast}F_{\alpha},
\end{align}
relativistic wave equation for electric and magnetic fields are derived and
its solution is given.Here we review that part shortly.Equation of motion in
dual field strength is written%
\begin{equation}
\partial_{\mu}\epsilon^{\mu\nu\gamma\ast}F_{\gamma}+\mu^{\ast}F^{\nu}=J^{%
\nu}.
\end{equation}
Multiply $-\epsilon_{\alpha\beta\nu}$ and taking trace we obtain 
\begin{equation}
\partial_{\mu}(g_{\alpha}^{\mu}g_{\beta}^{\gamma}-g_{\beta}^{\mu}g_{\alpha
}^{\gamma})^{\ast}F_{\gamma}-\frac{\mu}{2}\epsilon_{\alpha\beta\nu}%
\epsilon^{\nu ab}F_{ab}=-\epsilon_{\alpha\beta\nu}J^{\nu},
\end{equation}
and%
\begin{equation}
\partial_{\alpha}^{\ast}F_{\beta}-\partial_{\beta}^{\ast}F_{\alpha}-\mu
F_{\alpha\beta}=-\epsilon_{\alpha\beta\nu}J^{\nu}.
\end{equation}
Taking divergence of the above equation, when the current is conserved $%
\partial_{\alpha}^{\ast}F_{\alpha}=0$,we have%
\begin{equation}
\square^{\ast}F_{\beta}-\mu\partial_{\alpha}F_{\alpha\beta}=-\epsilon
_{\alpha\beta\nu}\partial_{\alpha}J_{\nu}.
\end{equation}
Using equation of motion 
\begin{align}
\partial_{\alpha}F_{\alpha\beta} & =-\frac{\mu}{2}\epsilon_{\beta\gamma
\delta}F^{\gamma\delta}+J_{\beta}  \notag \\
& =-\frac{\mu}{2}\epsilon_{\beta\gamma\delta}\epsilon^{\gamma\delta\alpha
\ast}F_{\alpha}+J_{\beta}  \notag \\
& =-\frac{\mu}{2}2\delta_{\beta}^{\alpha\ast}F_{\alpha}+J_{\beta}  \notag \\
& =-\mu^{\ast}F_{\beta}+J_{\beta},
\end{align}
twice,we get%
\begin{align}
\square^{\ast}F_{\beta} & =\mu\partial_{\alpha}F_{\alpha\beta}-\epsilon
_{\alpha\beta\nu}\partial_{\alpha}J_{\nu}  \notag \\
&
=\mu(-\mu^{\ast}F_{\beta}+J_{\beta})-\epsilon_{\alpha\beta\nu}\partial_{%
\alpha}J_{\nu}, \\
(\square+\mu^{2})^{\ast}F_{\mu} & =\mu(g_{\mu\nu}-\frac{\epsilon_{\mu
\alpha\nu}}{\mu}\partial_{\alpha})J_{\nu}.
\end{align}
Solution of dual field strength is given 
\begin{equation}
^{\ast}F_{\mu}=\frac{\mu J_{\mu}-\epsilon_{\mu\nu\alpha}\partial_{\nu
}J_{\alpha}}{\square+\mu^{2}},
\end{equation}
where 
\begin{align}
(\square+\mu^{2})\Delta(x) & =-i\delta^{3}(x), \\
\Delta(x) & =\frac{-i}{\square+\mu^{2}}\delta^{3}(x),  \notag \\
& =\int\frac{d^{3}k}{(2\pi)^{3}i}\frac{e^{-ik\cdot x}}{\mu^{2}-k^{2}+i%
\epsilon}=\frac{e^{-\mu\sqrt{r^{2}-t^{2}}}}{4\pi i\sqrt{r^{2}-t^{2}+i\epsilon%
}}.
\end{align}
In position space we have the solution of wave equation with source $J(y)$%
\begin{align}
^{\ast}F_{\mu}(x) & =\int d^{3}y\Delta(x-y)(\mu J_{\mu}(y)-\epsilon
_{\mu\alpha\nu}\partial_{\alpha}^{y}J_{\nu}(y)) \\
& =\int d^{3}y(\Delta(x-y)\mu
g_{\mu\nu}+\epsilon_{\mu\alpha\nu}(\partial_{\alpha}^{y}\Delta(x-y))J_{%
\nu}(y).
\end{align}
Here we neglect the homogeneous solution.We have shown here that both
electric and magnetic fields are massive and short range which decreases
exponentially $\exp(-\mu r)/4\pi r.$Let us return to the equation of
motion.Surface integral of first equation at large radius yields%
\begin{align}
\int\triangledown\cdot\mathbf{E}dS & \mathbf{=}\int E_{n}dr=2\pi RE_{n}, \\
2\pi RE_{n}-\mu\int BdS & \mathbf{=}\int\rho dS\mathbf{=}Q.
\end{align}
Radial component of electric field $E_{n}(R)$ decrease as $\exp(-\mu R)/4\pi
R$ for large $R$.We can safely neglect the first term at spatial
infinity.The $-\int drB$ ,is time-independent which follows from
conservation of $^{\ast }F^{\mu},$ and proportional to total charge.Static
solution of the vector potential is given for classical field%
\begin{align}
-\mu\int_{S}(\nabla\times\mathbf{A})\cdot d\mathbf{S} & \mathbf{=}\int
dSj^{0}(r)=Q, \\
-\mu\doint \mathbf{A\cdot dl} & \mathbf{=-}\mu\doint \mathbf{%
\nabla\theta\cdot dl=-}2\pi\mu\theta=Q,
\end{align}%
\begin{equation}
\mathbf{A}(x)_{|x|\rightarrow\infty}\rightarrow\frac{-Q}{2\pi\mu}\nabla
\arctan(\frac{y}{x}).
\end{equation}

That is even though magnetic field is short range, magnetic potential is
long range.Therefore we find that if QED action contains Chern-Simons term
and charged particle,we have a vortex solution for vector potential as an
Aharonov-Bohm effect.The above vortex solution with quantization of $\mu$ is
well known in quantum Hall systems which explains quantization of Hall
conductance and the possibility of statistics transmutaion.Hereafter we will
examine possibility of wash away condensate of chiral order parameter in QED$%
_{2+1}$ with massless four component fermion by the existence of vortex.

\subsection{ Chiral symmetry and Ward-Takahashi relation}

Here we consider the dynamical effects as mass generation in the presence of
Chern-Simons term. If the fermion is massless $m=0$,\tciLaplace\ has $U(2)$
symmetry generated by $\left\{ I,\gamma_{3},\gamma_{5},\tau\right\}
,\{\gamma_{\mu},\gamma_{\nu}\}=2g_{\mu\nu}$ for $\mu ,\nu=(0,1,2),$ $%
\gamma_{0}=\left( 
\begin{array}{cc}
\sigma_{3} & 0 \\ 
0 & -\sigma_{3}%
\end{array}
\right) ,\gamma_{1,2}=-i\left( 
\begin{array}{cc}
\sigma_{1,2} & 0 \\ 
0 & -\sigma_{1,2}%
\end{array}
\right) ,\gamma_{3}=\left( 
\begin{array}{cc}
0 & I \\ 
I & 0%
\end{array}
\right) ,\gamma_{5}=\left( 
\begin{array}{cc}
0 & -iI \\ 
iI & 0%
\end{array}
\right) ,\tau=-i[\gamma_{3},\gamma_{5}]/2=\left( 
\begin{array}{cc}
I & 0 \\ 
0 & -I%
\end{array}
\right) .\gamma_{3}$ and $\gamma_{5}$ act as chiral transformation%
\begin{align}
\psi(x) & \rightarrow e^{i\gamma_{3}\theta_{3}}\psi(x),  \notag \\
\psi(x) & \rightarrow e^{i\gamma_{5}\theta_{5}}\psi(x).
\end{align}
Scalar density is mixed by above transformation%
\begin{align}
\overline{\psi}(x)\psi(x) & \rightarrow\cos(2\theta_{3,5})\overline{\psi }%
(x)\psi(x)+i\sin(2\theta_{3,5})\overline{\psi}(x)\gamma_{3,5}\psi (x), 
\notag \\
\overline{\psi}(x)\tau\psi(x) & \rightarrow\overline{\psi}(x)\tau\psi(x).
\end{align}
Dynamical mass generation breaks $U(2)$ symmetry down to $U(1)_{I}\times
U(1)_{\tau}$ which is generated by $\{I,\tau\}$ ,\cite{BURDEN}.There exists
discrete parity transformation. 
\begin{equation}
P\psi(t,x,y)P^{-1}\rightarrow i\gamma^{1}\gamma^{3}\psi(t,-x,y).
\end{equation}
Parity even and odd mass are transformed 
\begin{align}
Pm_{e}\overline{\psi}\psi P^{-1} & \rightarrow m_{e}\overline{\psi}\psi, \\
Pm_{o}\overline{\psi}\tau\psi P^{-1} & \rightarrow-m_{o}\overline{\psi}%
\tau\psi.
\end{align}
Both of these mass terms are invariant under $U(1)_{I}\times U(1)_{\tau}$
transformation.%
\begin{align}
\psi(x) & \rightarrow e^{i\theta}\psi(x),  \notag \\
\psi(x) & \rightarrow e^{i\tau\theta_{\tau}}\psi(x).
\end{align}
We find the eigenvalue of the free particle Hamiltonian 
\begin{equation}
H=\gamma^{0}(\gamma^{i}p^{i}+m_{e}I+m_{o}\tau)
\end{equation}
as $E^{2}=p^{2}+m_{\pm}^{2},m_{\pm}=m_{e}\pm m_{o},$where $i=1,2,I$ is a $%
4\times4$ unit matrix and $\tau$ is a operator defined above.Two kinds of
mass are written by%
\begin{equation}
m_{e}I+m_{o}\tau=\left( 
\begin{array}{cc}
m_{+} & 0 \\ 
0 & m_{-}%
\end{array}
\right) =\left( 
\begin{array}{cc}
m_{e}+m_{o} & 0 \\ 
0 & m_{e}-m_{o}%
\end{array}
\right) .
\end{equation}
So that we may split 4-component spinor into upper and lower components by
projection operator%
\begin{equation}
\psi_{\pm}=\chi_{\pm}\psi=\chi_{\pm}\left( 
\begin{array}{c}
\psi_{+} \\ 
\psi_{-}%
\end{array}
\right) ,\chi_{\pm}=\frac{1\pm\tau}{2},\chi_{+}=\left( 
\begin{array}{cc}
1 & 0 \\ 
0 & 0%
\end{array}
\right) ,\chi_{-}=\left( 
\begin{array}{cc}
0 & 0 \\ 
0 & 1%
\end{array}
\right) .
\end{equation}
From the Lagrangian 
\begin{equation}
\tciLaplace=\overline{\psi}_{+}(i\gamma\cdot\partial-m_{+})\psi_{+}+%
\overline{\psi}_{-}(i\gamma\cdot\partial-m_{-})\psi_{-}
\end{equation}
we have the free propagator 
\begin{equation}
S_{0}(p)=-\frac{\gamma\cdot p+m_{+}}{p^{2}-m_{+}^{2}}\chi_{+}-\frac {%
\gamma\cdot p+m_{-}}{p^{2}-m_{-}^{2}}\chi_{-}.
\end{equation}
Since Chern-Simons term induces parity odd fermion mass,it is convenient to
introduce parity odd mass from the beginnings and choose the basis of the
eigenvalue of the Hamiltonian.Chiral representation is enough for this
purposes.In our 4-component spinor representation parity violating mass $%
\overline{\psi}\tau\psi$ is a pseudo scalar and current $J_{\mu}^{3,5}=%
\overline{\psi}\gamma_{3,5}\gamma_{\mu}\psi$ transform as an vector,which
mix each other under parity and charge conjugation transformation with
arbitrary phase.Starting from Dirac equation%
\begin{equation}
(\gamma\cdot\partial+m_{0})\psi=ie\gamma\cdot A\psi,\overline{\psi}%
(\gamma\cdot\overleftarrow{\partial}-m_{0})=-ie\overline{\psi}\gamma\cdot A,
\end{equation}
combination of above two equation,we obtain%
\begin{equation}
\partial_{\lambda}J_{\lambda}^{3,5}=2m_{0}J^{3,5}=D,
\end{equation}
where 
\begin{equation}
J_{\lambda}^{3,5}=\overline{\psi}\gamma_{3,5}\gamma_{\lambda}\psi ,J^{3,5}=%
\overline{\psi}\gamma_{3,5}\psi.
\end{equation}
A more general proof by Ward-Takahashi-relation of the vector current,%
\begin{align}
\partial_{\mu}^{x}T(J_{3,5\mu}(x)\psi(y)\overline{\psi}(z)) & =D+\delta
^{(3)}(x-y)\delta\psi(y)\overline{\psi}(z)+\delta^{(3)}(x-z)\psi (y)\delta%
\overline{\psi}(z)  \notag \\
&
=D+\gamma_{3,5}S_{F}(x-z)\delta^{(3)}(x-y)+\delta^{(3)}(x-z)S_{F}(y-z)%
\gamma_{3,5}
\end{align}
by operator of symmetry transformation, 
\begin{align}
\delta(x_{0}-y_{0})[J_{3,50}(x),\psi(y)] & =\delta\psi(y)\delta
^{(3)}(x-y)=\gamma_{3,5}\psi(y)\delta^{(3)}(x-y), \\
\delta(x_{0}-y_{0})[J_{3,50}(x),\overline{\psi}(y)] & =\delta\overline{\psi }%
(y)\delta^{(3)}(x-y)=\overline{\psi}(y)\gamma_{3,5}\delta^{(3)}(x-y).
\end{align}
In terms of the vertex function it is written%
\begin{equation}
(p-q)_{\mu}\Gamma_{3,5\mu}(p,q)=2m_{0}\Gamma_{3,5}(p,q)+%
\gamma_{3,5}S_{F}^{-1}(q)+S_{F}^{-1}(p)\gamma_{3,5}.
\end{equation}%
\begin{align*}
\Gamma_{3,5\mu}(p,q) & =\int d^{3}xd^{3}ye^{i(p\cdot y-q\cdot
x)}\left\langle TJ_{3,5\mu}(0)\psi(y)\overline{\psi}(x)\right\rangle _{T}, \\
\Gamma_{3,5}(p,q) & =\int d^{3}xd^{3}ye^{i(p\cdot y-q\cdot x)}\left\langle
TJ_{3,5}(0)\psi(y)\overline{\psi}(x)\right\rangle _{T},
\end{align*}
where the subscript $T$ means truncation of the fermion propagator.For the
proof of Ward-Takahashi relation in terms of integral equation for $%
\Gamma_{3,5},\Gamma_{3\mu,5\mu}$ see \cite{MASKAWA}.For vanishing bare mass $%
m_{0}$,current conservation is spontaneously broken by parity even mass
generation%
\begin{equation}
\lim_{p\rightarrow
q}(p-q)_{\mu}\Gamma_{3,5\mu}(p,q)=\{\gamma_{3,5},S_{F}^{-1}(p)\}\neq0.
\end{equation}
In other form it is well known 
\begin{equation}
\partial_{\mu}^{x}T(J_{5\mu}(x)\overline{\psi}(y)\gamma_{5}\psi (y))=-2%
\overline{\psi}(x)\psi(x).
\end{equation}
Taking vacuum expectation value of both sides of equation,non-vanishing of
the r.h.s indicates the existence of massless scalar NG boson,in the vertex $%
\Gamma_{3,5\mu}(p,q)$,of the form%
\begin{equation}
\frac{if_{\pi}(p-q)_{\mu}X_{\pi}(P=0,p-q)}{(p-q)^{2}},
\end{equation}
where $X_{\pi}$ is a Bethe-Salpeter amplitude for total momentum $P=0$.It is
equal to scalar part of propagator by Ward-Takahashi relation. If the right
hand side is finite and we have two order parameter $\left\langle \overline{%
\psi}\psi\right\rangle ,\left\langle \overline{\psi}\tau \psi\right\rangle $
of chiral symmetry breaking and parity violation. Dyson-Schwinger equation
is familiar to obtain non-perturbative propagator and dynamical
mass.Therefore we set up the Dyson-Schwinger equation for the electron
propagator in this representation.

\section{Dyson-Schwinger equation}

\subsection{quenched case}

Propagator of fermion and gauge boson are given 
\begin{align}
S(p) & =\frac{i}{A(p)\gamma\cdot p-B(p)}\rightarrow i\frac{(\gamma\cdot
pA_{+}(p)+B_{+}(p))}{p^{2}A_{+}^{2}(p)-B_{+}^{2}(p)}\chi_{+}+i\frac {%
(\gamma\cdot pA_{-}(p)+B_{-}(p))}{p^{2}A_{-}^{2}(p)-B_{-}^{2}(p)}\chi_{-}, \\
D_{\mu\nu}(k) & =\frac{1}{i}[\frac{g_{\mu\nu}-k_{\mu}k_{\nu}/(k^{2}+i%
\epsilon)-i\mu\epsilon_{\mu\nu\rho}k^{\rho}/(k^{2}+i\epsilon)}{k^{2}-\mu
^{2}+i\epsilon}+\xi\frac{k_{\mu}k_{\nu}}{(k^{2}+i\epsilon)^{2}}].
\end{align}
The quenched Schwinger-Dyson equation for the self-energy $\Sigma(p)$ is
written 
\begin{equation}
-i\Sigma(p)=(-ie)^{2}\int\frac{d^{3}q}{(2\pi)^{3}}\gamma_{\mu}S(q)\Gamma_{%
\nu }(p,q)D_{\mu\nu}(k),
\end{equation}
where $k=q-p.$In this normalization $\Sigma$ is real in the region $p^{2}<0.$%
We separate the vector and scalar part of the propagator by taking the trace
of $\Sigma.$Trace formulae for $(4\times4)$ $\gamma$ matrices are reduced to 
$(2\times2)$ ones with projection operator $\chi_{\pm}$%
\begin{align}
tr(\chi_{\pm}) & =2, \\
tr(\gamma_{\gamma}\gamma_{\nu}\chi_{\pm}) & =2g_{\mu\nu}, \\
tr(\gamma_{\mu}\gamma_{\nu}\gamma_{\rho}\chi_{\pm}) & =\mp2i\epsilon_{\mu
\nu\rho}.
\end{align}
In Chiral representation parity violating mass and vector part are different
sign for up and down component.We have the coupled integral equation for $%
A_{\pm}(p)$ and $B_{\pm}(p)$ in the Landau gauge by taking trace%
\begin{align}
i(S^{-1}-S_{0}^{-1}) & =((A(p)-1)\gamma\cdot p-B(p))=i\Sigma(p), \\
2(A(p)-1)p^{2} & =tr(i\gamma\cdot p\Sigma(p)),-2B(p)=tr(i\Sigma(p)).
\end{align}
To keep gauge invariance of the action we adopt the gauge covariant
approximation by introducing Ball-Chiu vertex ansatz for longitudinal
part[9].%
\begin{align}
\Gamma_{\mu}^{BC}(p,q) & =\Gamma_{\mu}^{T}(p,q)+\frac{A(p)+A(q)}{2}%
\gamma_{\mu}+\frac{A(p)-A(q)}{2(p^{2}-q^{2})}\gamma\cdot(p+q)(p+q)_{\mu } 
\notag \\
& -\frac{B(p)-B(q)}{p^{2}-q^{2}}(p+q)_{\mu}, \\
(p-q)_{\mu}\Gamma_{\mu}^{BC}(p,q) & =i(S^{-1}(q)-S^{-1}(p))  \notag \\
& =A(p)\gamma\cdot p-B(p)-(A(q)\gamma\cdot q-B(q)).
\end{align}
In this case the Dyson-Schwinger equations are following equations in the
Landau gauge 
\begin{equation}
B(p)_{\pm}=\frac{e^{2}}{4\pi^{2}}\int\frac{d^{3}q}{q^{2}A(q)_{%
\pm}^{2}+B(q)_{\pm}^{2}(k^{2}+\mu^{2})}[(A(p)_{\pm}+A(q)_{\pm})(B(q)_{\pm}%
\mp\mu A(q)_{\pm}(\frac{1}{2}-\frac{p^{2}-q^{2}}{2k^{2}}))  \notag
\end{equation}%
\begin{equation}
+\{\Delta A_{\pm}(B(q)_{\pm}\mp\frac{\mu}{2}A(q)_{\pm})-\Delta B_{\pm
}A(q)_{\pm}\}((p^{2}+q^{2})-\frac{k^{2}}{2}-\frac{(p^{2}-q^{2})^{2}}{2k^{2}}%
)],
\end{equation}%
\begin{align}
A(p)_{\pm} & =1+\frac{e^{2}}{4\pi^{2}p^{2}}\int\frac{d^{3}q}{q^{2}A(q)_{\pm
}^{2}+B(q)_{\pm}^{2}(k^{2}+\mu^{2})}[((A(p)_{\pm}+A(q)_{\pm})(\pm\mu
B(q)_{\pm}(\frac{1}{2}+\frac{p^{2}-q^{2}}{2k^{2}})  \notag \\
& +A(q)_{\pm}(\frac{(p^{2}-q^{2})^{2}}{4k^{2}}-\frac{k^{2}}{4}))+\{\Delta
A_{\pm}(\pm\frac{\mu}{2}B(q)_{\pm}-A(q)_{\pm}\frac{(p^{2}+q^{2})}{2})  \notag
\\
& -\Delta B_{\pm}(\mp\mu A(q)_{\pm}+B(q)_{\pm})\}((p^{2}+q^{2})-\frac{k^{2}}{%
2}-\frac{(p^{2}-q^{2})^{2}}{2k^{2}})]
\end{align}
where $k=q-p$ , 
\begin{equation}
\Delta A=\frac{A(p)-A(q)}{p^{2}-q^{2}},\Delta B=\frac{B(p)-B(q)}{p^{2}-q^{2}}%
.
\end{equation}
Angular integral formulae are given in Appendices A.

For bare vertex we set $\Gamma_{\mu}(p,q)=\gamma_{\mu}$. Recently we
obtained the numerical solutions with this approximation\cite{Raya},\cite%
{Hiroshima}.It is said that the vortex destroys condensate in
condensed matter physics.At bare vertex or inclusion the BC vertex we find $%
\mu_{cr}\simeq0.01e^{2}$ which is the same order of magnitude with Raya et.al%
\cite{Raya} in FIG.1.Simple vertex correction as $\gamma_{\mu}\rightarrow
A(p)\gamma_{\mu}$ yields $\mu_{cr}=8\cdot10^{-3}e^{2}.$
\begin{figure}
[ptb]
\begin{center}
\includegraphics[
height=2.7095in,
width=2.7095in
]%
{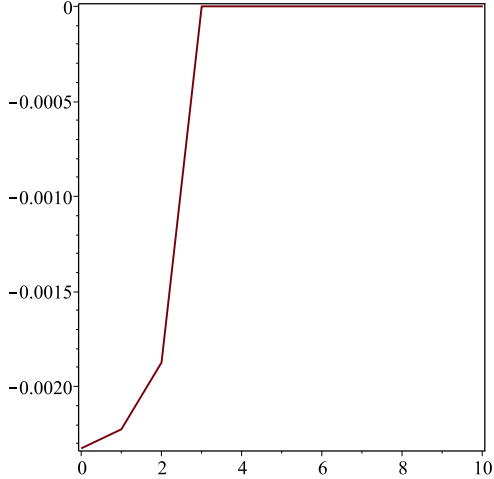}%
\caption{chiral order parameter $\left\langle \overline{\psi}\psi\right\rangle
$ for bare vertex with $\mu=10^{-2.3+0.1n}e^{2},n=0..7.$}%
\label{1}%
\end{center}
\end{figure}

 Numerically some instability arise from $\Delta BB(q)$ term in $A$
equation.This term is mass changing and compete the reduction of mass by
parity violating mass term.In pure QED this term is sensitive for low
momentum and infrared cut-off of momentum integration. The effects on vacuum
expectation value is relatively large.If we demand that the mass $B$
vanishes at the transition point and neglect this term,we find the clear
transition as the bare vertex case with the same critical point.We can
determine it by introducing renormalization of the topological mass as%
\begin{equation}
\mu_{R}=\mu+\frac{e^{2}}{2\pi}\theta(\mu-\mu_{cr}),
\end{equation}
where $\mu_{cr\text{ }}$is determined self-consistently in FIG.2.By this
analysis critical point $\mu_{cr}$ in the bare vertex case turned to be the
true one. Renormalization of topological mass is shown in Appendix.At least
for low energy we can safely neglect $B\Delta B$ terms and find clear phase
transition.In FIG.3 we see the parity violating order parameter $%
\left\langle \overline{\psi}\tau\psi\right\rangle $ as a function of
topological mass.Near the critical point $\mu_{cr}$,$\left\langle \overline{%
\psi}\tau\psi \right\rangle $ is lower than $\left\langle \overline{\psi}%
\psi\right\rangle $ which is shown in Fig.4 and first order phase transition
occurs.%

\begin{figure}
[ptb]
\begin{center}
\includegraphics[
height=2.7095in,
width=2.7095in
]%
{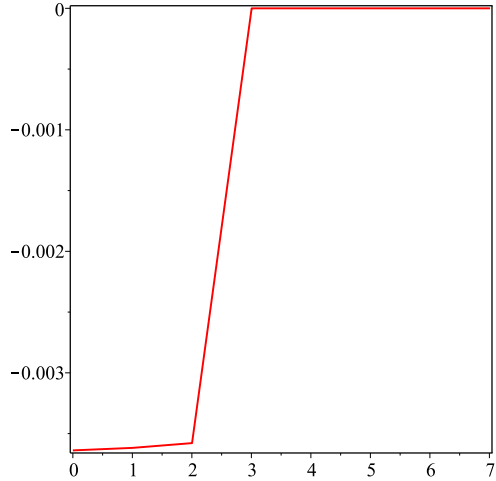}%
\caption{chiral order parameter $\left\langle \overline{\psi}\psi\right\rangle
 $ for $\mu=(10^{-2.3+0.1n}+\theta(n-3)/2\pi)e^{2}$ with BC vertex}%
\label{2}%
\end{center}
\end{figure}

\begin{figure}
[ptb]
\begin{center}
\includegraphics[
height=2.7095in,
width=2.7095in
]%
{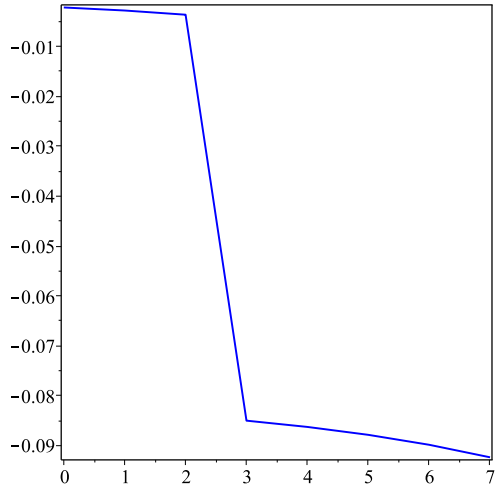}%
\caption{parity violating order parameter $\left\langle \overline{\psi}%
\tau\psi\right\rangle $ for $\mu=(10^{-2.3+0.1n}+\theta(n-3)/2\pi)e^{2}$ with
BC vertex.}%
\label{3}%
\end{center}
\end{figure}

\begin{figure}
[ptb]
\begin{center}
\includegraphics[
height=2.7095in,
width=2.7095in
]%
{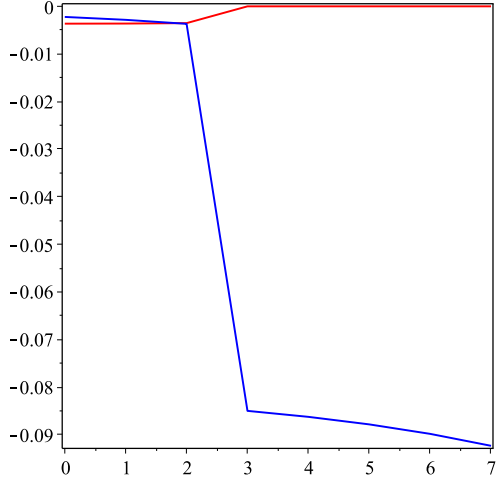}%
\caption{Parity even and odd order parameter $\left\langle \overline{\psi}%
\psi\right\rangle ,\left\langle \overline{\psi}\tau\psi\right\rangle $ for
$\mu=(10^{-2.3+0.1n}+\theta(n-3)/2\pi)e^{2}$ with BC vertex.}%
\label{4}%
\end{center}
\end{figure}
\begin{figure}
[ptb]
\begin{center}
\includegraphics[
height=2.7095in,
width=2.7095in
]%
{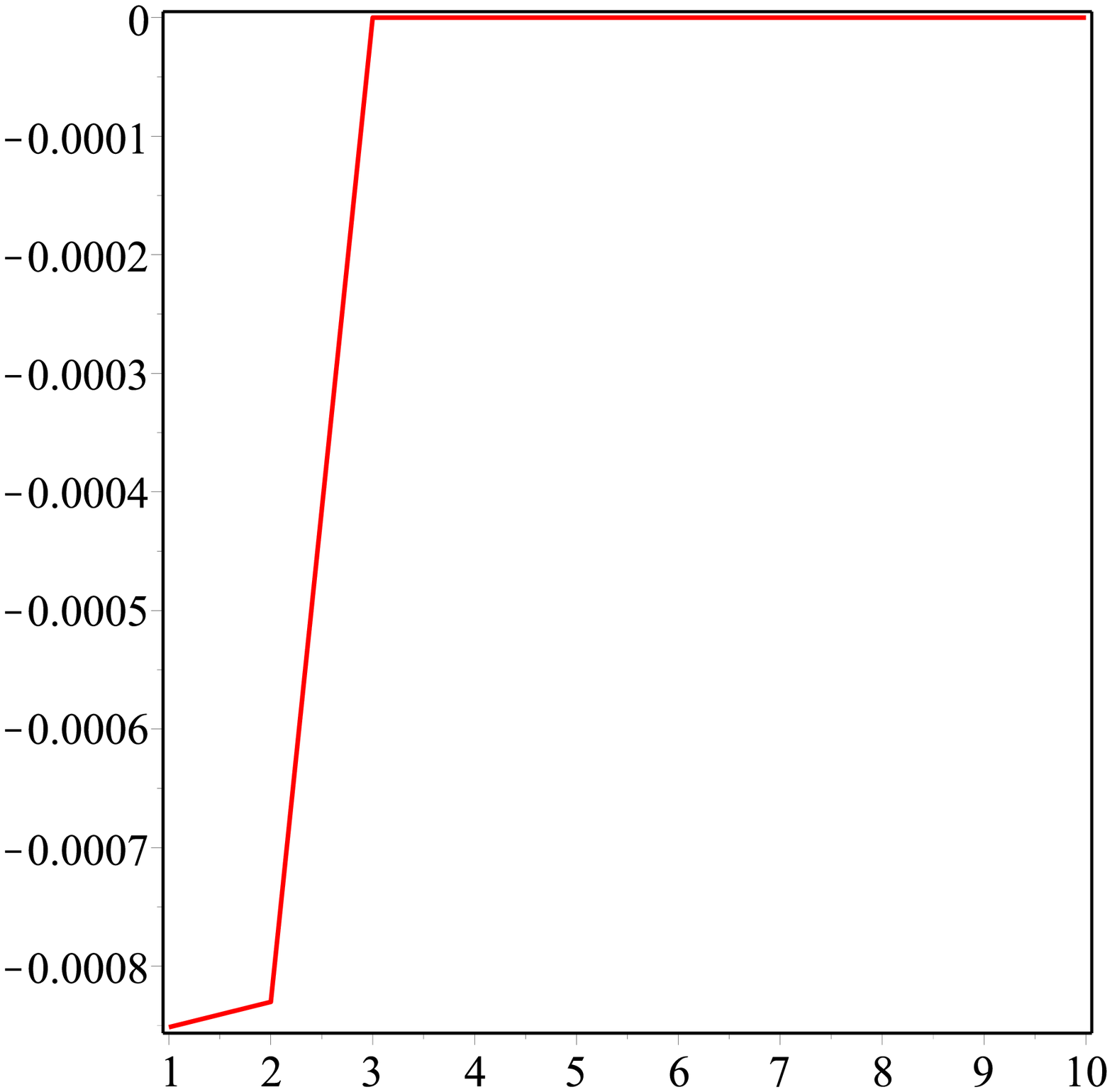}%
\caption{$\left\langle \overline{\psi}\psi\right\rangle $ for bare vertex with
one massless flavour loop correction.$\mu=(.008+.0005\cdot(n-1))e^{2}.$}%
\label{5}%
\end{center}
\end{figure}
\begin{figure}
[ptb]
\begin{center}
\includegraphics[
height=2.7095in,
width=2.7095in
]%
{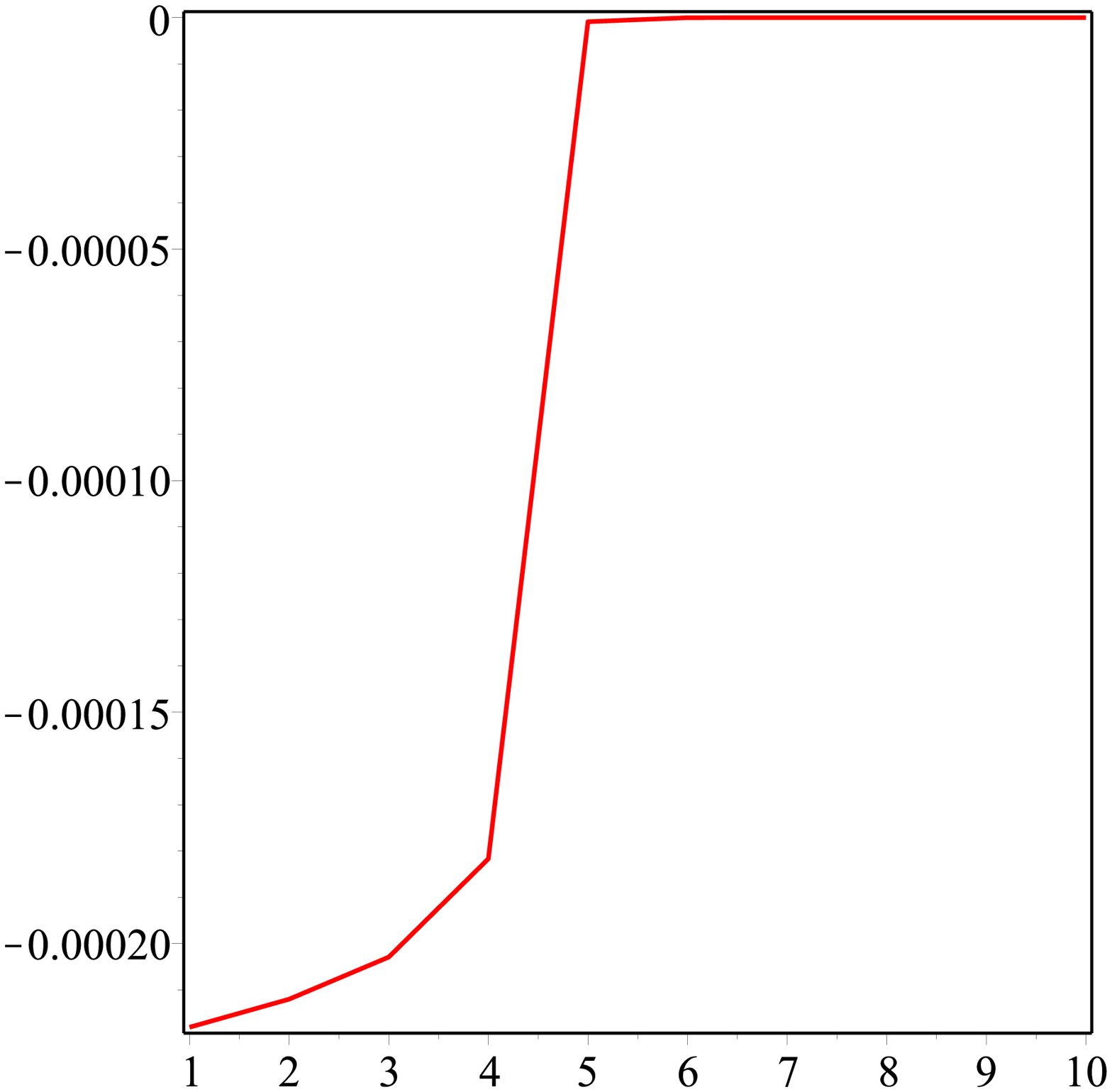}%
\caption{$\left\langle \overline{\psi}\psi\right\rangle $ for bare vertex with
two massless flavour loop correction with $\mu=(.005+.0005\cdot(n-1))e^{2}.$}%
\label{6}%
\end{center}
\end{figure}

\subsection{effects of vacuum polarization}

Using free vector propagator,

\begin{equation}
D_{\mu\nu}^{0}(p)=\frac{-i(g_{\mu\nu}-p_{\mu}p_{\nu}/p^{2}-i\mu\epsilon
_{\mu\nu\alpha}p^{\alpha}/p^{2})}{p^{2}-\mu^{2}+i\epsilon}-i\xi
p_{\mu}p_{\nu }/p^{4},
\end{equation}
we shall calculate one-loop corrections to the vector propagator%
\begin{equation}
(\mathit{D}^{\prime-1})_{\mu\nu}=(D_{0}^{-1})_{\mu\nu}-i\mathit{\Pi}%
_{\mu\nu},
\end{equation}
and set up the Dyson-Schwinger equation with bare vertex%
\begin{align}
\mathit{S}^{-1} & =S_{0}^{-1}+i\mathit{\Sigma,}  \notag \\
\mathit{\Sigma(p)} & =-ie^{2}\int\frac{d^{3}k}{(2\pi)^{3}}\gamma^{\mu
}S(p+k)\gamma^{\nu}D_{\mu\nu}^{^{\prime}}(k).
\end{align}
One-loop vacuum polarization is evaluated for fermion mass $m_{e}\neq
0,m_{o}=0$ with Pauli-Villars or dimensional reguralization to remove gauge
nonivariant cut-off dependent term and Parity-odd Chern-Simons term is not
induced here.See appendix and \cite{DJT} .%
\begin{align}
\Pi_{\mu\nu}(k) & \equiv ie^{2}N\int\overline{d^{3}}pTr(\gamma_{\mu}\frac {1%
}{\gamma\cdot p-m}\gamma_{\nu}\frac{1}{\gamma\cdot(p-k)-m})  \notag \\
& =-e^{2}N\frac{T_{\mu\nu}}{8\pi}[(\sqrt{k^{2}}+\frac{4m^{2}}{\sqrt{k^{2}}}%
)\ln(\frac{2m+\sqrt{k^{2}}}{2m-\sqrt{k^{2}}})-4m], \\
T_{\mu\nu} & =(g_{\mu\nu}-\frac{k_{\mu}k_{\nu}}{k^{2}}),\overline{d^{3}}p=%
\frac{d^{3}p}{(2\pi)^{3}},
\end{align}
Polarization function $\Pi(k)$ is 
\begin{align}
\Pi(k) & =\frac{-e^{2}}{8\pi}N[(\sqrt{k^{2}}+\frac{4m^{2}}{\sqrt{k^{2}}})\ln(%
\frac{2m+\sqrt{k^{2}}}{2m-\sqrt{k^{2}}})-4m],  \notag \\
& =\frac{-e^{2}}{8}N\sqrt{-k^{2}}(k^{2}>0,m=0),  \notag \\
& =\frac{-e^{2}N}{6\pi m}k^{2}+O(k^{4})(k^{2}/m\ll1).
\end{align}
For $m=0$ case 
\begin{equation}
\mathit{\Pi}_{\mu\nu}(k)=-\frac{e^{2}N}{8}(g_{\mu\nu}-\frac{k_{\mu}k_{\nu}}{%
k^{2}})\sqrt{-k^{2}},
\end{equation}
gives a non-perturbative correction to the vector propagator for low energy
in the chiral symmetric phase.From the relation%
\begin{equation}
(D^{-1})_{\mu\nu}D_{\nu\rho}=g_{\mu\rho},
\end{equation}
we have the inverse of the bare photon propagator%
\begin{equation}
(D^{-1})_{\mu\nu}=i(p^{2}g_{\mu\nu}-p_{\mu}p_{\nu}+i\mu\epsilon_{\mu\nu%
\alpha }p^{\alpha})+i\frac{p_{\mu}p_{\nu}}{\xi}.
\end{equation}
Adding the vacuum polarization of massless loop 
\begin{align}
(D^{\prime-1})_{\mu\nu} & =i(p^{2}g_{\mu\nu}-p_{\mu}p_{\nu}+i\mu\epsilon
_{\mu\nu\alpha}p^{\alpha})-i(-\frac{e^{2}N}{8}\sqrt{-p^{2}})(g_{\mu\nu}-%
\frac{p_{\mu}p_{\nu}}{p^{2}})+i\frac{p_{\mu}p_{\nu}}{\xi}  \notag \\
& =i[(p^{2}+\frac{e^{2}N}{8}\sqrt{-p^{2}})(g_{\mu\nu}-\frac{p_{\mu}p_{\nu}}{%
p^{2}})+i\mu\epsilon_{\mu\nu\alpha}p^{\alpha}]+i\frac{p_{\mu}p_{\nu}}{\xi},
\end{align}
we obtain the full propagator 
\begin{equation}
D^{\prime}{}_{\mu\nu}=-i\frac{(p^{2}-\pi(p^{2}))(g_{\mu\nu}-p_{\mu}p_{\nu
}/p^{2})-i\mu\epsilon_{\mu\nu\rho}p^{\rho}}{(p^{2}-\pi(p^{2}))^{2}-\mu
^{2}p^{2}}-i\xi\frac{p_{\mu}p_{\nu}}{p^{4}}.
\end{equation}
In Euclid space it has the form%
\begin{equation}
D_{\mu\nu}^{\prime}=-i\frac{A(-p^{2})(\delta_{\mu\nu}-p_{\mu}p_{\nu}/p^{2})-%
\mu\epsilon_{\mu\nu\rho}p^{\rho}}{A(-p^{2})^{2}+\mu^{2}p^{2}}-i\xi\frac{%
p_{\mu}p_{\nu}}{p^{4}},
\end{equation}
where 
\begin{equation}
A(-p^{2})=(p^{2}+\frac{e^{2}N}{8}\sqrt{p^{2}}).
\end{equation}
Including vacuum polarization for photon the Dyson-Schwinger equation has
the following form 
\begin{equation}
\Sigma(p^{2})=e^{2}\int\frac{d^{3}q}{(2\pi)^{3}}\gamma_{\mu}S(q)\gamma_{\nu
}D_{\mu\nu}^{\prime}(k).
\end{equation}
In Euclid space with $\xi=0$ gauge ,we have%
\begin{align}
B_{\pm}(p) & =\frac{e^{2}}{4\pi^{3}}\int d^{3}q[\frac{(k^{2}+\pi
(k^{2}))B_{\pm}(q)}{[A_{\pm}^{2}(q)q^{2}+B_{\pm}^{2}(q)](k^{2}+%
\pi(k^{2}))^{2}+\mu^{2}k^{2})}  \notag \\
& \mp\frac{\mu A_{\pm}(q)(q\cdot k)}{[A_{\pm}^{2}(q)q^{2}+B_{%
\pm}^{2}(q)](k^{2}+\pi(k^{2}))^{2}+\mu^{2}k^{2})}], \\
A_{\pm}(p) & =1+\frac{e^{2}}{4\pi^{3}p^{2}}\int d^{3}q[\frac{(k^{2}+\pi
(k^{2}))A_{\pm}(q)(p\cdot k)(q\cdot k)}{[A_{\pm}^{2}(q)q^{2}+B_{%
\pm}^{2}(q)](k^{2}+\pi(k^{2}))^{2}+\mu^{2}k^{2})k^{2}}  \notag \\
& \mp\frac{\mu B_{\pm}(q)(p\cdot k)}{[A_{\pm}^{2}(q)q^{2}+B_{%
\pm}^{2}(q)](k^{2}+\pi(k^{2}))^{2}+\mu^{2}k^{2})}].
\end{align}
Here we apply the Ball-Chiu vertex as in the quenched case.Kondo and Maris
applied $1/N$ approximation to solve the equation.However in the gauge
covariant approximation which satisfy Ward-Takahashi relation has been shown
that chiral order parameter in quenched and unquenched case are same at $N=1$
for weak coupling for $\mu=0$ case.So we choose covariant gauge and weak
coupling.It is easy to take BC vertex and improve the Dyson-Schwinger
equation as in the quenched case.But we have not enough memories to run
PC.So that we only show the massless loop correction in the Landau gauge
case. In this case critical value $\mu_{cr}$ in unquenched case is
approximately the same value for quenched case.For example we find
numerically $\mu_{cr}\sim0.01e^{2}$ for $N=1$ and $\mu_{cr}\sim0.008e^{2}$
for $N=2.$At the critical point we find that $m_{e}=m_{o}$, $B_{+}(p)=0$ and 
$m_{e}$ vanishes above the critical point.This is the destruction mechanism
of superfluidity by vortex in our model.
For $N\geq2$ case we have very small values of order parameter for small
topological mass.For large topological mass order parameter changes its sign
at some value of topological mass.There may be a strong coupling phase for $%
N\geq2.4$ at least for small topological mass,where vacuum expectation value 
$\left\langle \overline{\psi}\psi\right\rangle $ vanishes.For these cases $%
1/N$ expansion may be a good way to study the phase structure for strong
coupling region in our model.A famous critical number of flavour which was
derived in the linearized Schwinger-Dyson equation has been know as $%
N_{c}=32/\pi^{2}$ above which the chiral symmetry is restored.The results of 
$1/N$ expansion and the phase structure derived by KI.Kondo and P.Maris may
be realized\cite{Kondo}.

\section{Analysis by spectral function}

\subsection{definition of spectral function}

In this section we would like to determine critical value of topological
mass above which chiral condensate is washed away theoretically.First we
notice that the spectral function for the propagator as one of the
possibility\ cite{OPS}.In three dimension,absence of ultraviolet divergences
is important.If we know only infrared behaviour or the leading logarithm of
infrared divergence near the mass shell,it is possible to determine the
whole region of the propagator in position space by the anomalous
dimension,which is supplied by lowest ordered spectral function.By this
method we find that only short distance behaviour of the propagator is
modified and we have a finite chiral condensate for pure QED$_{2+1}$ \cite%
{HOSHINO}.If we choose soft-photon exponentiation to include all orders of
soft-photon emission by electron,its spectral function may be written as $%
e^{F}$, where $F$ is a model independent spectral function of the lowest
order in the coupling constant for pure QED$_{2+1}$ 
\begin{align}
\rho(x) & =e^{F(\mu|x|)}, \\
S_{F}(x) & =S_{F}^{0}(x)\rho(x).
\end{align}
Here we consider the fermion spectral function.The vacuum expectation value
of the anticommutator has the form \cite{BJF}

\begin{align}
iS^{\prime}(x,y) & =\left\langle 0|\{\psi(x),\overline{\psi}%
(y)\}|0\right\rangle  \notag \\
& =\sum_{n}[\left\langle 0|\psi(0)|n\right\rangle \left\langle n|\overline {%
\psi}(0)|0\right\rangle e^{-ip_{n}\cdot(x-y)}+\left\langle 0|\overline{\psi }%
(0)|n\right\rangle \left\langle n|\psi(0)|0\right\rangle e^{ip_{n}\cdot
(x-y)}].
\end{align}
We introduce the spectral amplitude by grouping together in the sum over $n$
all states of given three-momentum $q$%
\begin{equation}
\rho_{\alpha\beta}(q)=(2\pi)^{2}\sum_{n}\delta^{(3)}(p_{n}-q)\left\langle
0|\psi_{\alpha}(0)|n\right\rangle \left\langle n|\overline{\psi}_{\beta
}(0)|0\right\rangle
\end{equation}
and set out to construct its general form from invariance arguments.$\rho(q)$
is a $4\times4$ matrix and may be expanded in terms of $16$ linearly
independent products of $\gamma$ matrices.Under the assumptions of Lorentz
invariance and Parity transformation it reduces to the form%
\begin{equation}
\rho(q)_{\alpha\beta}=\rho_{1}(q)\gamma\cdot
q+\rho_{2}(q)\delta_{\alpha\beta }.
\end{equation}
Second term in (82) can be related directly to (83) with the aid of PCT
invariance of the vacuum \cite{BJF},\cite{BJQ}.Parity,Charge conjugation and
Time\ reversal transformation are defined in our representation of $\gamma$
matrices 
\begin{align}
P\psi(t,x,y)P^{-1} & =i\gamma^{1}\gamma^{3}\psi(t,-x,y), \\
PA^{0}(t,x,y)P^{-1} & =A^{0}(t,-x,y), \\
PA^{1}(t,x,y)P^{-1} & =-A^{1}(t,-x,y), \\
PA^{2}(t,x,y)P^{-1} & =A^{2}(t,-x,y),
\end{align}%
\begin{equation}
P\overline{\psi}(t,x,y)\gamma^{1}\psi(t,x,y)P^{-1}=-\overline{\psi }%
(t,-x,y)\gamma^{1}\psi(t,-x,y)
\end{equation}%
\begin{align}
C\psi(x)C^{-1} & =C\gamma^{0}\psi^{\ast}=C\overline{\psi}^{T}=\psi_{c}, \\
C & =\gamma^{2},C^{-1}\gamma^{\mu}C=-\gamma^{\mu T}, \\
CA^{\mu}(t,x,y)C^{-1} & =-A^{\mu}(t,x,y),
\end{align}%
\begin{align}
T\psi_{\alpha}(t,x,y)T^{-1} &
=T_{\alpha\beta}\psi_{\beta}(-t,x,y),T=-i\gamma^{2}\gamma^{3}, \\
TA^{0}(t,x,y)T^{-1} & =A^{0}(-t,x,y), \\
TA^{1,2}(t,x,y)T^{-1} & =-A^{1,2}(-t,x,y).
\end{align}
Effects of PCT transformation on $\overline{\psi}(y)\psi(x)$ is%
\begin{align}
PCT\psi_{\alpha}^{A}(t,x,y)T^{-1}C^{-1}P^{-1} & =-\gamma_{\gamma\alpha}^{1}{}%
\overline{\psi}_{\gamma}^{T}(-t,-x,y), \\
PCT\overline{\psi}_{\beta}^{B}(t,x,y)T^{-1}C^{-1}P^{-1} & =-\gamma
_{\lambda\beta}^{1}\psi_{\lambda}^{T}(-t,-x,y), \\
PCT\overline{\psi}_{\alpha}^{A}(t^{\prime},x^{\prime},y^{\prime})\psi_{\beta
}^{B}(t,x,y)T^{-1}C^{-1}P^{-1} &
=\gamma_{\alpha\gamma}^{1}\psi_{\gamma}^{B}(-t,-x,y)\overline{\psi}%
_{\lambda}^{A}(-t^{\prime},-x^{\prime},y^{\prime })\gamma_{\lambda\beta}^{1}.
\end{align}
Inserting (98) along with (83) into (82) and using $\gamma^{2T}=-\gamma^{2}$%
,we obtain finally

\begin{align}
iS_{\alpha\beta}^{\prime}(x-y) & =\int\frac{d^{3}q}{(2\pi)^{3}}\theta
(q_{0})([\gamma\cdot q\rho_{1}(q^{2})+\rho_{2}(q^{2})]_{\alpha\beta
}e^{-iq\cdot(x-y)}  \notag \\
& +\{\gamma^{1}[\gamma\cdot q\rho_{1}(q^{2})+\rho_{2}(q^{2})]\gamma
^{1}\}_{\alpha\beta}e^{iq\cdot(x^{\prime}-y^{\prime})})  \notag \\
& =\int\frac{d^{3}q}{(2\pi)^{2}}\theta(q_{0})[\rho_{1}(q^{2})i\gamma
\cdot\partial_{x}+\rho_{2}(q^{2})]_{\alpha\beta}(e^{-iq\cdot(x-y)}-e^{iq%
\cdot(x^{\prime}-y^{\prime})})  \notag \\
& =\int\frac{d^{3}q}{(2\pi)^{2}}[\theta(q_{0})\gamma\cdot
q\rho_{1}(q^{2})+\epsilon(q_{0})\rho_{2}(q^{2})]_{\alpha\beta}e^{-iq%
\cdot(x-y)},
\end{align}
where $x^{\prime}=(-t,-x_{1},x_{2}),(\gamma^{1})^{2}=-1.$Since $\rho$
vanishes for space-like $q^{2}$,we may also write this as an integral over
mass spectrum by introducing%
\begin{equation}
\rho(q^{2})=\int_{0}^{\infty}\rho(s)\delta(q^{2}-s)ds.
\end{equation}
We find 
\begin{align}
iS^{\prime}(x-y) & =-\int ds[i\rho_{1}(s)\gamma\cdot\partial+\rho
_{2}(s)]i\Delta(x-y;\sqrt{s})  \notag \\
& =\int ds\{\rho_{1}(s)iS(x-y;\sqrt{s})+[\sqrt{s}\rho_{1}(s)-\rho
_{2}(s)]i\Delta(x-y;\sqrt{s})\}
\end{align}
where invariant $\Delta$ function is given%
\begin{align}
i\Delta^{\prime}(x,y) & \equiv\left\langle 0|[\phi(x),\phi
(y)]|0\right\rangle \\
& =\sum_{n}\left\langle 0|\phi(0)|n\right\rangle \left\langle n|\phi
(0)|0\right\rangle (e^{-iP_{n}\cdot(x-y)}-e^{iP_{n}\cdot(x-y)})  \notag \\
& =\frac{1}{(2\pi)^{2}}\int d^{3}q\rho(q^{2})\theta(q_{0})(e^{-iq\cdot
(x-y)}-e^{iq\cdot(x-y)}) \\
& =\frac{1}{(2\pi)^{2}}\int_{0}^{\infty}ds\rho(s)\int d^{3}q\delta
(q^{2}-s)\epsilon(q_{0})e^{-iq\cdot(x-y)}  \notag \\
& =\int_{0}^{\infty}ds\rho(s)i\Delta(x-y,\sqrt{s}).
\end{align}
The above spectral representation goes through unchanged for the vacuum
expectation value of the time-ordered product of Dirac field;it is necessary
only to replace the $iS$ and $i\Delta$ by the Feynman propagator $S_{F}$ and 
$\Delta_{F}$.If we know the matrix element $\left\langle 0|\psi_{\alpha
}(0)|n\right\rangle $,we can determine the spectral function $\rho_{1}$and $%
\rho_{2}$.Perturbative $O(e^{2})$ spectral function can be obtained by the
usual definition 
\begin{align}
\rho^{(2)}(p^{2}) & =\int\frac{d^{3}xe^{-ip\cdot x}}{(2\pi)^{3}}\int \frac{%
d^{2}r}{(2\pi)^{2}}e^{ir\cdot x}\frac{d^{2}k}{(2\pi)^{2}}e^{ik\cdot
x}\left\langle 0|\psi(0)|r,k\right\rangle \left\langle r,k|\overline{\psi }%
(0)|0\right\rangle  \notag \\
& =\int\frac{d^{3}xe^{-ip\cdot x}}{(2\pi)^{3}}\frac{d^{2}r}{(2\pi)^{2}}%
e^{ir\cdot x}\frac{d^{2}k}{(2\pi)^{2}}e^{ik\cdot x}\sum_{\lambda,S}T_{1}%
\overline{T_{1}}.
\end{align}
In perturbation theory one photon emission matrix element is given%
\begin{align}
T_{1} & =\left\langle 0|\psi(0)|r,k\right\rangle \simeq\left\langle
0^{in}|T[U(\infty,-\infty)\psi^{in}(0)]r;k\text{ in}\right\rangle  \notag \\
& =-i\left\langle 0^{in}|T[\psi^{in}(0),e\int d^{3}x\overline{\psi}%
^{in}(x)\gamma_{\mu}\psi^{in}(x)A_{\mu}^{in}(x)]|r;k\text{ in}\right\rangle
\\
& =-ie\int d^{3}xS_{F}^{0}(0-x)\gamma_{\mu}\left\langle
0|\psi^{in}(x)|r\right\rangle \left\langle 0|A_{\mu}^{in}(x)|k\right\rangle 
\notag \\
& =\frac{-ie}{\gamma\cdot(r+k)-m+i\epsilon}\gamma_{\mu}\epsilon_{\lambda
}^{\mu}(k)U_{S}(r)\sqrt{\frac{m}{E_{r}}}\frac{1}{\sqrt{2k_{0}}}.
\end{align}
For the evaluation of $\rho(p^{2})$,if we integrate $x$ first,we obtain $%
\delta^{(3)}(k+r-p)$ for energy-momentum conservation.In our case first we
integrate $k$.After that we exponentiate the function $F$ and integrate $r$
in the non perturbative case.At that stage the results are position
dependent.Finally we obtain the spectral function in position space $\rho(x)$
for infinite number of photon emission%
\begin{align}
\rho(x) & =\int\frac{d^{2}re^{ir\cdot x}}{(2\pi)^{2}}  \notag \\
& \times\sum_{n=0}^{\infty}\frac{1}{n!}(\int\frac{d^{2}k}{(2\pi)^{2}}%
\theta(k_{0})\delta(k^{2})e^{ik\cdot
x}\sum_{\lambda})_{n}\delta^{(3)}(p-r-\dsum \limits_{i=1}^{\infty}k_{i})T_{n}%
\overline{T}_{n},
\end{align}
where the notation $f(k)_{0}=1,$ $f(k)_{n}=\dprod \limits_{i=1}^{n}f(k_{i})$
is used and $T_{n}\overline{T}_{n}$ may be replaced to $(T_{1}\overline{T}%
_{1})^{n}$ in our approximation.The polarization sum for gauge boson is
given 
\begin{equation}
\sum_{\lambda}\epsilon_{\lambda}^{\mu}(k)\epsilon_{\lambda}^{\nu}(k)=-[g^{%
\mu\nu}-(1-\xi)\frac{k^{\mu}k^{\nu}}{k^{2}}-i\mu\epsilon^{\mu\nu\rho }\frac{%
k_{\rho}}{k^{2}}].
\end{equation}
Since we do not fix the numbers of photon $n$ we sum up from $n\ $equals
zero to infinity.The function 
\begin{equation}
F=\int\frac{d^{2}k}{(2\pi)^{2}}\sum\limits_{\lambda,S}T_{1}\overline{T_{1}}%
e^{ik\cdot x},
\end{equation}
is a phase space integral of one photon intermediate state which may be
exponentiated as $e^{F}.$We use the trace formula to get 
\begin{align}
\sum\limits_{\lambda,S}T_{1}\overline{T_{1}} & =\gamma\cdot rA(r)+B(r). \\
A(r) & =\frac{1}{4m^{2}}Tr(r\cdot\gamma\sum\limits_{\lambda,S}T_{1}\overline{%
T_{1}}),  \notag \\
B(r) & =\frac{1}{4}Tr(\sum\limits_{\lambda,S}T_{1}\overline{T_{1}}).
\end{align}
To order $e^{2}$ we have model independent spectral function for 4-component
fermion 
\begin{equation}
F(\mu|x|)=-e^{2}\int\frac{d^{3}k}{(2\pi)^{2}}e^{-ik\cdot x}\theta(k_{0})%
\frac{\gamma\cdot p+m}{2m}[\delta(k^{2}-\mu^{2})(\frac{m^{2}}{(r\cdot k)^{2}}%
+\frac{1}{r\cdot k})-(\xi-1)\frac{\partial}{\partial k^{2}}%
\delta(k^{2}-\mu^{2})].
\end{equation}
Hereafter we keep in mind the projection operator of positive energy $%
(\gamma\cdot p+m)/2m$ for the propagator.From this form $\rho_{2}=m\rho_{1}$
in our approximation.To obtain the explicit form of the function $F$, we use
parameter trick 
\begin{align}
\lim_{\epsilon\rightarrow0}\int_{0}^{\infty}d\alpha e^{-\alpha(\epsilon
-ik\cdot r)} & =\frac{i}{k\cdot r}, \\
\lim_{\epsilon\rightarrow0}\int_{0}^{\infty}\alpha d\alpha e^{-\alpha
(\epsilon-ik\cdot r)} & =-\frac{1}{(k\cdot r)^{2}},
\end{align}
combined with positive frequency part of the propagator with bare mass $\mu$%
\begin{align}
D^{+}(x) & \equiv\int\frac{d^{3}k}{i(2\pi)^{2}}\theta(k^{0})\delta(k^{2}-%
\mu^{2})e^{ik\cdot x}  \notag \\
& =\frac{1}{i(2\pi)^{2}}\int_{0}^{\infty}\frac{2\pi kdkJ_{0}(k\left\vert
x\right\vert )}{2\sqrt{k^{2}+\mu^{2}}}=\frac{e^{-\mu|x|}}{4\pi i\left\vert
x\right\vert },
\end{align}
we obtain the function $F$ symbolically as 
\begin{align}
F & =ie^{2}m^{2}\int_{0}^{\infty}\alpha d\alpha D^{+}(x+\alpha
r)-e^{2}\int_{0}^{\infty}d\alpha D^{+}(x+\alpha r)-ie^{2}(\xi-1)\frac{%
\partial }{\partial\mu^{2}}D^{+}(x,\mu^{2})  \notag \\
& =ie^{2}m^{2}F_{1}(x)-e^{2}F_{2}(x)+e^{2}(\xi-1)F_{L}.
\end{align}

We get each terms 
\begin{align}
F_{2} & =\int\frac{d^{3}k}{(2\pi)^{2}}\theta(k^{0})\delta(k^{2}-\mu
^{2})e^{ik\cdot x}\frac{1}{r\cdot k}  \notag \\
& =\lim_{\epsilon\rightarrow0}\int_{0}^{\infty}d\alpha e^{-\alpha
(\epsilon-ik\cdot r)}\int\frac{d^{3}k}{i(2\pi)^{2}}\theta(k^{0})\delta
(k^{2}-\mu^{2})e^{ik\cdot x}  \notag \\
& =\int_{0}^{\infty}d\alpha D^{+}(x+\alpha r)=\frac{E_{1}(\mu|x|)}{4\pi m},
\\
F_{1} & =-\int\frac{d^{3}k}{(2\pi)^{2}}\theta(k^{0})\delta(k^{2}-\mu
^{2})e^{ik\cdot x}\frac{1}{(r\cdot k)^{2}}  \notag \\
& =\lim_{\epsilon\rightarrow0}\int_{0}^{\infty}\alpha d\alpha e^{-\alpha
(\epsilon-ik\cdot r)}\frac{d^{3}k}{(2\pi)^{2}}\theta(k^{0})\delta(k^{2}-%
\mu^{2})e^{ik\cdot x}  \notag \\
& =\int_{0}^{\infty}\alpha d\alpha D^{+}(x+\alpha r)=\frac{\exp(-\mu
|x|)-\mu|x|E_{1}(\mu|x|)}{4\pi m^{2}\mu i}, \\
F_{L} & =-i\frac{\partial}{\partial\mu^{2}}\frac{e^{-\mu|x|}}{4\pi
i\left\vert x\right\vert }=-\frac{1}{8\pi\mu}\frac{\partial}{\partial\mu }(%
\frac{e^{-\mu|x|}}{|x|})  \notag \\
& =\frac{1}{8\pi}\frac{\exp(-\mu|x|)}{\mu},
\end{align}%
\begin{align}
E_{1}(z) & =\int_{z}^{\infty}\frac{e^{-t}}{t}dt,(|\arg z|<\pi) \\
E_{1}(\mu|x|) & \sim-\gamma-\ln(\mu|x|),\mu|x|\ll1,\text{ } \\
E_{1}(\mu|x|) & \sim\frac{\exp(-\mu|x|)}{\mu|x|}(1-\frac{1}{\mu|x|}+\frac {2%
}{(\mu|x|)^{2}})(\mu|x|\gg1),
\end{align}
where $r^{2}=m^{2}$. For short distance we have 
\begin{equation}
F\simeq\frac{e^{2}}{4\pi m}(\gamma+\ln(\mu|x|))+\frac{(\xi+1)e^{2}}{8\pi}%
\frac{1}{\mu}.
\end{equation}
Above spectral function contains linear infrared divergent term proportional
to $1/\mu.$This term depends on the gauge parameter.So we choose $\xi=-1$
gauge to drop it.Other terms are independent of $\xi.$In 4-dimension
well-known infrared behaviour of charged particle is reproduced in this
way.In QED$_{3}$ we have used this technique to determine a short-distance
behaviour of the propagator.In the long distance region propagator behaves
as free one for finite $\mu$.If we set anomalous dimension which is a
coefficient of $\ln(\mu|x|)$ to be unity $e^{2}/4\pi m=1,$we obtain the
propagator at short distance as%
\begin{align}
S_{F}(x) & =-\frac{(i\gamma\cdot\partial+m)e^{-m|x|}(\mu|x|)e^{\gamma}}{%
4\pi|x|}  \notag \\
& =-(i\gamma\cdot\partial+m)\frac{\mu e^{\gamma}}{4\pi}e^{-m|x|},
\end{align}
which shows condensation $-iTr(S_{F}(x))=finite$.In this case the trace
means positive energy part only.For massless gauge boson vacuum expectation
value is infrared cut-off dependent.If we set $\mu=m=e^{2}/4\pi,$we have $%
\left\langle \overline{\psi}\psi\right\rangle =e^{4}/32\pi^{2}$ which is
very close to the numerical value by solving Dyson-Schwinger equation in
Euclid momentum space.In our topologically massive QED,there are two order
parameters $\left\langle \overline{\psi}\psi\right\rangle ,\left\langle 
\overline{\psi }\tau\psi\right\rangle $ that have non vanishing value for
small topological mass.

\subsection{contribution of Chern-Simons term}

Just above the critical point there exists only parity violating mass with
non-vanishing order parameter $\left\langle \overline{\psi}\tau\psi
\right\rangle $.So that we may expect that the chiral symmetry breaking mass
vanishes and parity violating condensate remains at the critical point.We
show this is the case in evaluating the lowest order spectral function with
Chern-Simons term.In the case of parity violation,we may include parity
violating part of the spectral function $\overline{\rho}\tau+\overline{\rho }%
_{\mu}(\gamma_{\mu}\tau)$.Since we adopt the chiral representation of the
propagator in solving the Dyson-Schwinger equation,we will determined
spectral functions in chiral representation too. Contribution of
Chern-Simons term for the spectral function in chiral representation is
given in the Appendix B%
\begin{equation}
F_{CS}^{+}=-e^{2}\int\frac{d^{3}k}{(2\pi)^{2}}e^{ik\cdot
x}\theta(k_{0})\tau\lbrack\gamma\cdot r\{(\frac{-1}{\mu}+\frac{\mu}{m^{2}})%
\frac{1}{8r\cdot k}-\frac{\mu}{8(r\cdot k)^{2}}\}+\frac{\mu}{m}\frac{1}{%
8r\cdot k}+\frac{m\mu }{4(r\cdot k)^{2}})]\delta(k^{2}-\mu^{2}).
\end{equation}
After the integration we obtain the correction of wave function
renormalization by Chern-Simons term ($1/r\cdot k$ term) as $F_{2}$ 
\begin{equation}
F^{+}(\mu|x|)=(-\frac{\gamma\cdot r+m}{2m}\frac{e^{2}}{8\pi m}+\frac {%
\gamma\cdot r}{m}\frac{e^{2}}{32\pi\mu}-\frac{e^{2}\mu}{32\pi m^{2}}%
)E_{1}(\mu|x|),
\end{equation}
for $S_{+}$.At short distance we take into account only $E_{1}(\mu
|x|)\sim-\gamma-\ln(\mu|x|)$.If we exponentiate $F(\mu|x|)$ as $e^{F}$ using 
$(\gamma\cdot r)^{2}/m^{2}=1,$have 
\begin{align}
e^{A\gamma\cdot r/m\ln(\mu|x|)} & =\cosh(A\ln(\mu|x|))+\frac{\gamma\cdot r}{m%
}\sinh(A\ln(\mu|x|))  \notag \\
& =\frac{\gamma\cdot r+m}{2m}(\mu|x|)^{A}+\frac{m-\gamma\cdot r}{2m}%
(\mu|x|)^{-A}.
\end{align}
Final form is written as 
\begin{equation}
e^{F}=\frac{\gamma\cdot r+m}{2m}(\mu|x|)^{A}\exp(\gamma+\frac{e^{2}\mu}{%
32\pi m^{2}}\ln(\mu|x|)),A=\frac{e^{2}}{8\pi m}-\frac{e^{2}}{32\pi\mu}, 
\notag
\end{equation}
for short distance.In chiral representation if $A=1,$we have non vanishing
condensation of $\left\langle \overline{\psi}\psi\right\rangle _{+}$%
.Otherwise $\left\langle \overline{\psi}\psi\right\rangle _{+}=0$.Below the
critical value of the topological mass $\mu\leq\mu_{cr}$ there are two kind
of mass with different anomalous dimension.In that case we cannot separate
them in the chiral representation.However at the $\mu=\mu_{cr}$ there is
only parity odd mass and and its condensate.So that we set wave function
renormalization $A=0$,other contribution is only logarithmic.Therefore we
set $e^{2}/8\pi m=1$ and have $e^{2}/32\pi\mu=1$ for critical value of $\mu$%
.In this case we have only parity violating order parameter $\left\langle 
\overline{\psi}\tau \psi\right\rangle $ above the critical point.This is
just the desirable form to give parity odd order parameter $\left\langle 
\overline{\psi}\tau \psi\right\rangle .$So that at the critical point chiral
order parameter vanishes and it undergoes into parity violating phase. So we
conclude that the critical value of the topological mass is $%
\mu_{cr}=e^{2}/32\pi\simeq .10^{-2}e^{2}$ which is totally consistent with
the value in our numerical analysis of Dyson-Schwinger equation for quenched
case.This transition is the same as Kosterlitz-Thouless type at finite
temperature where single vortex excitation destroys the superfluidity.So
that the Topological Massive QED$_{3}$ is the same with QED$_{3}$ with
single vortex at zero temperature except for small topological mass.In the
statistical model,behaviour of the propagator near the critical point is
studied with renormalization group equation for vortex number (chemical
potential) and the inverse temperature[2,3]. In our model spectral function
provide us anomalous dimension and we can determine the ultraviolet
behaviour and critical point in the existence of topological
mass.Fortunately our Dyson-Schwinger equation successfully determines the
structure of the propagator and dynamical mass near the critical region.

\section{\protect\bigskip Summary}

In this work we studied the dynamics of Kosterlitz-Thouless type transition
in Topologically Massive Abelian Gauge Theory in three dimensional
space-time.In this model equation of motion contain vortex solution for the
vector potential.We showed that there exists a critical value of the
topological mass above which chiral condensate is washed away for four
component fermion.This phenomenon turned out to be gauge invariant by the
choice of BC vertex in solving Dyson-Schwinger equation for fermion
self-energy.In the analysis of spectral function we showed it modify the
short distance behaviour of the propagator in position space.In that case
anomalous dimension control the order parameter of chiral condensate in the
absence of topological mass.However if we add Chern-Simons term to the
Lagrangian parity odd part of the gauge boson propagator is destructive to
parity even part. In this way lowest order spectral function vanishes at the
critical value of topological mass and the chiral condensate is washed away.
Our next step is to evaluate critical temperature by solving Dyson-Schwinger
equation or spectral function,temperature dependence of specific heat and
compare them with the experiment.

\section{Appendices}

\subsection{angular integral}

\subsubsection{quenced case}

\begin{align}
I_{0}(p,q) & =\int_{-1}^{1}\frac{d\cos\theta}{k^{2}+\mu^{2}}=\frac{-1}{2pq}%
\ln(\frac{(p-q)^{2}+\mu^{2}}{(p+q)^{2}+\mu^{2}}), \\
I_{1}(p,q) & =\int_{-1}^{1}\frac{d\cos\theta}{k^{2}+\mu^{2}}((p^{2}+q^{2})-%
\frac{k^{2}}{2}-\frac{(p^{2}-q^{2})^{2}}{2k^{2}}), \\
I_{2}(p,q)_{\pm} & =\int_{-1}^{1}\frac{d\cos\theta}{k^{2}+\mu^{2}}(\frac {1}{%
2}\pm\frac{p^{2}-q^{2}}{2k^{2}})  \notag \\
& =\frac{-1}{4pq}\ln(\frac{(p-q)^{2}+\mu^{2}}{(p+q)^{2}+\mu^{2}})\pm \frac{%
p^{2}-q^{2}}{4\mu^{2}pq}\ln(\frac{1+\mu^{2}/(p-q)^{2}}{1+\mu ^{2}/(p+q)^{2}}%
), \\
I_{3}(p,q) & =\int_{-1}^{1}\frac{d\cos\theta}{k^{2}+\mu^{2}}(\frac {%
(p^{2}-q^{2})^{2}}{4k^{2}}-\frac{k^{2}}{4})  \notag \\
& =\frac{(p^{2}-q^{2})^{2}}{8\mu^{2}pq}\ln(\frac{1+\mu^{2}/(p-q)^{2}}{%
1+\mu^{2}/(p+q)^{2}})-\frac{1}{2}-\frac{\mu^{2}}{8pq}\ln(\frac{%
(p-q)^{2}+\mu^{2}}{(p+q)^{2}+\mu^{2}}).
\end{align}
After angular integral we have the following coupled integral equation 
\begin{align}
B(p)_{\pm} & =\frac{e^{2}}{4\pi^{2}}\int_{0}^{\infty}\frac{dqq^{2}}{%
q^{2}A(q)_{\pm}^{2}+B(q)_{\pm}^{2}}[(A(p)_{\pm}+A(q)_{\pm})(B(q)_{%
\pm}I_{0}[p,q]\mp\mu A(q)_{\pm}I_{2}(p,q)_{-})  \notag \\
& +\{\Delta A_{\pm}(B(q)_{\pm}\mp\frac{\mu}{2}A(q)_{\pm})-\Delta B_{\pm
}A(q)_{\pm}\}I_{1}(p,q)],
\end{align}%
\begin{align}
A(p)_{\pm} & =1+\frac{e^{2}}{4\pi^{2}p^{2}}\int_{0}^{\infty}\frac{dqq^{2}}{%
q^{2}A(q)_{\pm}^{2}+B(q)_{\pm}^{2}}[((A(p)_{\pm}+A(q)_{\pm})(\pm\mu
B(q)_{\pm}I_{2}(p,q)_{+}+A(q)_{\pm}I_{3}(p,q))  \notag \\
& +\{\Delta A_{\pm}(\pm\frac{\mu}{2}B(q)_{\pm}-A(q)_{\pm}\frac{(p^{2}+q^{2})%
}{2})-\Delta B_{\pm}(\mp\mu A(q)_{\pm}+B(q)_{\pm})\}I_{1}(p,q)].
\end{align}

\subsubsection{unquenched case}

For unquenched case, to evaluate angular integral we may use complex number
to represent parity even and odd piece of the photon propagator 
\begin{align}
\func{Re}(\frac{1}{k^{2}+\pi(k^{2})+i\mu\sqrt{k^{2}}}) & =\frac{%
k^{2}+\pi(k^{2})}{(k^{2}+\pi(k^{2}))^{2}+\mu^{2}k^{2}}, \\
\func{Im}(\frac{1}{k^{2}+\pi(k^{2})+i\mu\sqrt{k^{2}}}) & =\frac {-\mu\sqrt{%
k^{2}}}{(k^{2}+\pi(k^{2}))^{2}+\mu^{2}k^{2}}.
\end{align}
It has been known that the integral kernel is a logarithmic function as%
\begin{equation}
K(p,q)\propto\ln(\frac{|p-q|+c}{p+q+c})
\end{equation}
where $c=e^{2}N/8.$ So that the angular integral with finite topological
mass $\mu$ may be a analytic continuation from the case $\mu=0$ to $\mu\neq0.
$%
\begin{equation}
K(p,q)\propto\ln(\frac{|p-q|+c+i\mu}{p+q+c+i\mu}).
\end{equation}
For example%
\begin{equation}
J_{0}(p,q)=\int_{-1}^{1}\frac{dt}{(p^{2}+q^{2}-2pqt)+(c+i\mu)\sqrt{%
p^{2}+q^{2}-2pqt}}=\frac{1}{pq}\ln(\frac{p+q+c+i\mu}{|p-q|+c+i\mu}).
\end{equation}
The integration of the type 
\begin{equation}
-\mu\int_{-1}^{1}\frac{d\cos\theta\text{ }q\cdot k}{(k^{2}+ck)^{2}+\mu
^{2}k^{2}}=\func{Im}\int_{-1}^{1}\frac{d\cos\theta\text{ }q\cdot (q-p)}{%
((p-q)^{2}+(c+i\mu)\sqrt{(p-q)^{2})}\sqrt{(p-q)^{2}}},
\end{equation}
is rewritten 
\begin{align}
J_{2}(p,q)_{-} & =-\int_{-1}^{1}\func{Im}(\frac{dt}{(p^{2}+q^{2}-2pqt)+(c+i%
\mu)\sqrt{p^{2}+q^{2}-2pqt}})\frac{(p^{2}-q^{2})-(p-q)^{2}}{2\sqrt{%
p^{2}+q^{2}-pqt}}  \notag \\
& =\frac{-1}{2pq}\func{Im}(p+q-|p-q|-(c+i\mu)\ln(\frac{p+q+c+i\mu }{%
|p-q|+c+i\mu}))  \notag \\
& -\frac{p^{2}-q^{2}}{2pq}\func{Im}(\frac{1}{c+i\mu}\ln (\frac{%
(p+q)(|p-q|+c+i\mu)}{|p-q|(p+q+c+i\mu)}).
\end{align}
In the same way 
\begin{equation}
-\mu\int_{-1}^{1}\frac{d\cos\theta\text{ }p\cdot k}{(k^{2}+ck)^{2}+\mu
^{2}k^{2}}=\func{Im}\int_{-1}^{1}\frac{d\cos\theta\text{ }p\cdot (q-p)}{%
((p-q)^{2}+(c+i\mu)\sqrt{(p-q)^{2})}\sqrt{(p-q)^{2}}},
\end{equation}
$p\cdot(q-p)=-(p^{2}-q^{2})/2-(p-q)^{2}/2$ leads%
\begin{align}
J_{2}(p,q)_{+} & =-\int_{-1}^{1}\func{Im}(\frac{dt}{(p^{2}+q^{2}-2pqt)+(c+i%
\mu)\sqrt{p^{2}+q^{2}-2pqt}})\frac{(p-q)^{2}+(p^{2}-q^{2})}{2\sqrt{%
p^{2}+q^{2}-pqt}}  \notag \\
& =\frac{-1}{2pq}\func{Im}(p+q-|p-q|-(c+i\mu)\ln(\frac{p+q+c+i\mu }{%
|p-q|+c+i\mu}))  \notag \\
& +\frac{p^{2}-q^{2}}{2pq}\func{Im}(\frac{1}{c+i\mu}\ln (\frac{%
(p+q)(|p-q|+c+i\mu)}{|p-q|(p+q+c+i\mu)}).
\end{align}
Since%
\begin{equation*}
\frac{(p\cdot k)(q\cdot k)}{k^{2}}=-\frac{(p-q)^{2}}{4}+\frac{%
(p^{2}-q^{2})^{2}}{4(p-q)^{2}}
\end{equation*}%
\begin{align}
J_{3}(p,q) & =\func{Re}\int_{-1}^{1}\frac{d\cos\theta}{(p-q)^{2}+(c+i\mu)%
\sqrt{(p-q)^{2}}}\frac{(p\cdot k)(q\cdot k)}{k^{2}}  \notag \\
& =-\frac{1}{4pq}\func{Re}[2pq+(c+i\mu)(|p-q|-(p+q))+(c+i\mu)^{2}\ln(\frac{%
p+q+c+i\mu}{|p-q|+c+i\mu})  \notag \\
& \frac{-(p+q)^{2}|p-q|+(p-q)^{2}(p+q)}{(c+i\mu)}+\frac{(p^{2}-q^{2})^{2}}{%
(c+i\mu)^{2}}\ln(\frac{(p+q)(|p-q|+c+i\mu)}{|p-q|(p+q+c+i\mu)})].
\end{align}
The Dyson-Schwinger equations are rewritten as 
\begin{align}
B_{\pm}(p) & =\frac{e^{2}}{2\pi^{2}}\int q^{2}dq[\frac{B_{\pm}(q)J_{0}(p,q)}{%
[A_{\pm}^{2}(q)q^{2}+B_{\pm}^{2}(q)]}\pm\frac{A_{\pm}(q)J_{2}(p,q)_{-}}{%
[A_{\pm}^{2}(q)q^{2}+B_{\pm}^{2}(q)]}],  \notag \\
A_{\pm}(p) & =1+\frac{e^{2}}{2\pi^{2}p^{2}}\int q^{2}dq[\frac{A_{\pm
}(q)J_{3}(p,q)}{[A_{\pm}^{2}(q)q^{2}+B_{\pm}^{2}(q)]}\pm\frac{%
B_{\pm}(q)J_{2}(p,q)_{+}}{[A_{\pm}^{2}(q)q^{2}+B_{\pm}^{2}(q)]}].
\end{align}

\subsection{ \protect\bigskip evaluation of vacuum polarization}

In a two-dimensional representation, the trace of products of four $\gamma $%
-matrices are: 
\begin{align*}
tr(I_{2}) & =2, \\
tr(\gamma^{\mu}) & =0, \\
tr(\gamma^{\mu}\gamma^{\nu}) & =2g^{\mu\nu}, \\
tr(\gamma^{\mu}\gamma^{\nu}\gamma^{\rho}) & =-i\epsilon^{\mu\nu\rho},
\end{align*}%
\begin{equation}
tr(\gamma^{\mu}\gamma^{\nu}\gamma^{\rho}\gamma^{\sigma})=2(g^{\mu\nu}g^{\rho%
\sigma}-g^{\mu\rho}g^{\nu\sigma}+g^{\mu\sigma}g^{\nu\rho}).
\end{equation}%
\begin{equation}
\Pi_{\mu\nu}(k)\equiv-e^{2}\int\frac{d^{3}p}{(2\pi)^{3}i}Tr(\gamma_{\mu}%
\frac{1}{\gamma\cdot p-m}\gamma_{\nu}\frac{1}{\gamma\cdot(p-k)-m}).
\end{equation}
Substituting the chiral representation of the propagator into $\Pi_{\mu\nu
}(k)$%
\begin{equation}
S(p)=\frac{i}{\gamma\cdot p-m+i\epsilon}\rightarrow i(\frac{\gamma\cdot
p+m_{+}}{p^{2}-m_{+}^{2}})\chi_{+}+i(\frac{\gamma\cdot p+m_{-}}{%
p^{2}-m_{-}^{2}})\chi_{-},
\end{equation}%
\begin{align}
& Tr(\gamma_{\mu}(\frac{\gamma\cdot p+m_{+}}{p^{2}-m_{+}^{2}}\chi_{+}+\frac{%
\gamma\cdot p+m_{-}}{p^{2}-m_{-}^{2}}\chi_{-})\gamma_{\nu}(\frac {%
\gamma\cdot(p-k)+m_{+}}{(p-k)^{2}-m_{+}^{2}}\chi_{+}+\frac{\gamma
\cdot(p-k)+m_{-}}{(p-k)^{2}-m_{-}^{2}}\chi_{-}))  \notag \\
& =Tr(\gamma_{\mu}\frac{\gamma\cdot p+m_{+}}{p^{2}-m_{+}^{2}}%
\chi_{+}\gamma_{\nu}\frac{\gamma\cdot(p-k)+m_{+}}{(p-k)^{2}-m_{+}^{2}}\chi
_{+})+Tr(\gamma_{\mu}\frac{\gamma\cdot p+m_{-}}{p^{2}-m_{-}^{2}}%
\chi_{-}\gamma_{\nu}\frac{\gamma\cdot(p-k)+m_{-}}{(p-k)^{2}-m_{-}^{2}}%
\chi_{-}),
\end{align}%
\begin{align}
& Tr(\gamma_{\mu}(\gamma\cdot p+m_{+})\chi_{+}\gamma_{\nu}(\gamma
\cdot(p-k)+m_{+})\chi_{+})  \notag \\
&
=2(p_{\mu}(p-k)_{\nu}+p_{\nu}(p-k)_{\mu}-g_{\mu\nu}(p%
\cdot(p-k)-m_{+}^{2})+im_{+}\epsilon_{\mu\nu\rho}p^{\rho}-im_{+}\epsilon_{%
\mu\nu\rho }(p-k)^{\rho}).
\end{align}%
\begin{align}
& Tr(\gamma_{\mu}(\gamma\cdot p+m_{-})\chi_{-}\gamma_{\nu}(\gamma
\cdot(p-k)+m_{-})\chi_{-})  \notag \\
&
=2(p_{\mu}(p-k)_{\nu}+p_{\nu}(p-k)_{\mu}-g_{\mu\nu}(p%
\cdot(p-k)-m_{-}^{2})-im_{-}\epsilon_{\mu\nu\rho}p^{\rho}+im_{-}\epsilon_{%
\mu\nu\rho }(p-k)^{\rho}).
\end{align}%
\begin{equation}
\Pi_{\mu\nu}(k)=-e^{2}\int_{0}^{1}dx\dint \frac{d^{3}p^{\prime}}{(2\pi)^{3}i}%
\frac{2N_{\mu\nu}}{(-p^{^{\prime}2}+K)^{2}},
\end{equation}
where%
\begin{align}
K & =-k^{2}x(1-x)-m^{2},p^{\prime}=p-k(1-x), \\
N_{\mu\nu} & =[-m^{2}+x(1-x)k^{2}]\delta_{\mu\nu}-\frac{1}{3}p^{\prime
2}\delta_{\mu\nu}-2k_{\mu}k_{\nu}x(1-x).
\end{align}
Regulating the ultraviolet divergence by cut-off 
\begin{equation}
\int_{0}^{\Lambda}\frac{p^{\prime4}dp^{\prime}}{(-p^{\prime2}+K)^{2}i}=-%
\frac{3}{2}\arctan(\frac{\Lambda}{K})K+\Lambda+O(\frac{1}{\Lambda}),\int%
\frac{d^{3}p}{(2\pi)^{3}}\frac{1}{(p^{2}+K)^{2}}=\frac{1}{8\pi}\frac {1}{%
\sqrt{K}}.
\end{equation}
we obtain the vacuum polarization tensor for two-component fermion%
\begin{align}
\Pi_{\mu\nu} & (-k^{2})=-e^{2}\int_{0}^{1}dx\int\frac{d^{3}p^{\prime}}{%
(2\pi)^{3}}\frac{2N_{\mu\nu}}{(p^{\prime2}+k^{2}x(1-x)+m^{2})^{2}}  \notag \\
& =\frac{-4e^{2}}{8\pi}(\delta_{\mu\nu}k^{2}-k_{\mu}k_{\nu})\int_{0}^{1}dx%
\frac{x(1-x)}{\sqrt{m^{2}+k^{2}x(1-x)}}+\frac{e^{2}\Lambda}{3\pi^{2}}%
\delta_{\mu\nu} \\
& =\frac{-e^{2}}{8\pi}(\delta_{\mu\nu}-\frac{k_{\mu}k_{\nu}}{k^{2}})[(\sqrt{%
-k^{2}}+\frac{4m^{2}}{\sqrt{-k^{2}}})\arctan(\frac{\sqrt{-k^{2}}}{2m})+2m]+%
\frac{e^{2}\Lambda}{3\pi^{2}}\delta_{\mu\nu}.
\end{align}
In Minkowski space we have%
\begin{equation}
\Pi_{\mu\nu}(k^{2})=\frac{e^{2}}{16\pi}(g_{\mu\nu}-\frac{k_{\mu}k_{\nu}}{%
k^{2}})((\sqrt{k^{2}}+\frac{4m^{2}}{\sqrt{k^{2}}})\ln(\frac{2m-\sqrt{k^{2}}}{%
2m+\sqrt{k^{2}}})+4m)-\frac{e^{2}\Lambda}{3\pi^{2}}g_{\mu\nu}.
\end{equation}
From the above expression, vacuum polarization tensor for chiral
representation of fermion is given 
\begin{equation*}
\Pi_{\mu\nu}(p)=-\frac{2e^{2}}{3\pi^{2}}\Lambda g_{\mu\nu}+T_{\mu\nu}\frac{%
e^{2}}{8\pi}p^{2}(\int_{2|m_{+}|}^{\infty}\frac{da(1+4m_{+}^{2}/a^{2})}{%
p^{2}-a^{2}+i\epsilon}+\int_{2|m_{-}|}^{\infty}\frac{da(1+4m_{-}^{2}/a^{2})}{%
p^{2}-a^{2}+i\epsilon}) 
\end{equation*}%
\begin{equation}
+i\epsilon_{\mu\nu\rho}p^{\rho}\frac{e^{2}}{2\pi}(m_{+}\int_{2|m_{+}|}^{%
\infty}\frac{da}{p^{2}-a^{2}+i\epsilon}-m_{-}\int_{2|m_{-}|}^{\infty}\frac{da%
}{p^{2}-a^{2}+i\epsilon}).
\end{equation}
subsection{\bigskip spectral function}

Here we derive the contributions of Chern-Simons term to the spectral
function.For scalar and vector part are given by trace%
\begin{equation}
B(r,k)_{+}=\frac{e^{2}}{16m(r\cdot k)^{2}}tr((r+k)\cdot\gamma+m)\chi_{+}%
\gamma_{\mu}(r\cdot\gamma+m)\gamma_{\nu}((r+k)\cdot\gamma+m))i\mu\epsilon
_{\mu\nu\rho}\frac{k_{\rho}}{\mu^{2}},
\end{equation}%
\begin{equation}
A(r,k)_{+}=\frac{e^{2}}{16m^{3}(r\cdot k)^{2}}tr(r\cdot\gamma((r+k)\cdot
\gamma+m)\chi_{+}\gamma_{\mu}(r\cdot\gamma+m)\gamma_{\nu}((r+k)\cdot
\gamma+m))i\mu\epsilon_{\mu\nu\rho}\frac{k_{\rho}}{\mu^{2}},
\end{equation}
where $\gamma$ matrices are $4\times4.$First we evaluate $B(r,k).$%
\begin{align}
& tr(((r+k)\cdot\gamma+m)^{2}\chi_{+}\gamma_{\mu}(r\cdot\gamma+m)\gamma_{\nu
}i\epsilon_{\mu\nu\rho}\frac{k_{\rho}}{\mu}  \notag \\
& =tr([2(m^{2}+r\cdot
k)+k^{2}+2m(r+k)\cdot\gamma]\chi_{+}\gamma_{\mu}(r\cdot\gamma+m)\gamma_{%
\nu})i\epsilon_{\mu\nu\rho}\frac{k_{\rho}}{\mu }  \notag \\
& =[(2(m^{2}+r\cdot k)+k^{2})r_{\sigma}tr(\chi_{+}\gamma_{\mu}\gamma_{\sigma
}\gamma_{\nu})+2m^{2}(r+k)_{\sigma}tr(\chi_{+}\gamma_{\sigma}\gamma_{\mu
}\gamma_{\nu})]i\epsilon_{\mu\nu\rho}\frac{k_{\rho}}{\mu}.
\end{align}%
\begin{align}
B(r,k) & =-\frac{4(2(m^{2}+r\cdot k)+\mu^{2})r\cdot k}{32m(r\cdot k)^{2}\mu }%
+\frac{8m^{2}(r\cdot k+\mu^{2})}{32m(r\cdot k)^{2}\mu}  \notag \\
& =-\frac{4\mu r\cdot k}{32m(r\cdot k)^{2}}-\frac{1}{4m\mu}+\frac{8m^{2}\mu 
}{32m(r\cdot k)^{2}}.  \notag \\
& =\frac{\mu}{8m(r\cdot k)}-\frac{1}{4m\mu}+\frac{\mu m}{4(r\cdot k)^{2}}.
\end{align}
Next we evaluate $A(r,k)$%
\begin{align}
& tr(r\cdot\gamma((r+k)\cdot\gamma+m)\chi_{+}\gamma_{\mu}(r\cdot
\gamma+m)\gamma_{\nu}((r+k)\cdot\gamma+m))  \notag \\
& =tr((r+k)\cdot\gamma+m)r\cdot\gamma((r+k)\cdot\gamma+m)\chi_{+}\gamma_{\mu
}(r\cdot\gamma+m)\gamma_{\nu})  \notag \\
& =tr((r+k)\cdot\gamma+m)(m^{2}+r\cdot k+mr\cdot\gamma)\chi_{+}\gamma_{\mu
}(r\cdot\gamma+m)\gamma_{\nu})  \notag \\
& =tr((2m(m^{2}+r\cdot k)+(2m^{2}+r\cdot k)r\cdot\gamma+(m^{2}+r\cdot
k)k\cdot\gamma)\chi_{+}\gamma_{\mu}(r\cdot\gamma+m)\gamma_{\nu})  \notag \\
& =2m(m^{2}+r\cdot
k)r_{\sigma}tr(\chi_{+}\gamma_{\mu}\gamma_{\sigma}\gamma_{\nu})+m(2m^{2}+r%
\cdot
k)r_{\sigma}tr(\chi_{+}\gamma_{\sigma}\gamma_{\mu}\gamma_{\nu})+m(m^{2}+r%
\cdot k)k_{\sigma}tr(\chi_{+}\gamma _{\sigma}\gamma_{\mu}\gamma_{\nu}) 
\notag \\
& =4mi(m^{2}+r\cdot k)r_{\sigma}\epsilon_{\sigma\mu\nu}-2mi(2m^{2}+r\cdot
k)r_{\sigma}\epsilon_{\sigma\mu\nu}-2mi(m^{2}+r\cdot
k)k_{\sigma}\epsilon_{\sigma\mu\nu}.
\end{align}%
\begin{align}
A(r,k) & =\frac{1}{32m^{3}(r\cdot k)^{2}}tr(r\cdot\gamma((r+k)\cdot
\gamma+m)\gamma_{\mu}(r\cdot\gamma+m)\gamma_{\nu}((r+k)\cdot\gamma
+m))i\epsilon_{\mu\nu\rho}\frac{k_{\rho}}{\mu}  \notag \\
& =\frac{1}{32m^{3}(r\cdot k)^{2}}(-8m(m^{2}+r\cdot k)\frac{r\cdot k}{\mu }%
+4m(2m^{2}+r\cdot k)\frac{r\cdot k}{\mu}+4m(m^{2}+r\cdot k)\mu)  \notag \\
& =-\frac{1}{8m^{2}\mu}-\frac{\mu}{8(r\cdot k)^{2}}-\frac{1}{8\mu(r\cdot k)}+%
\frac{\mu}{8m^{2}(r\cdot k)}.
\end{align}

\end{document}